\documentclass{aa}
\usepackage{epsfig}
\def \eg {{\it e.g.} }
\def \etal {{\it et al.} }
\def \ie {{\it i.e. } }
\newcommand{\new}[1]{{ { #1}}}
\newcommand{\nouveau}[1]{{ { #1}}}
\newcommand{\toutnouveau}[1]{{ {#1}}}
\newcommand{\no}[1]{{ {#1}}}
\newcommand{\neuf}[1]{{{#1}}}
\def \tomega {\tilde\omega}
\begin{document}

\title{Propagation of warps in moderately thick disks} 

\author{F.~Masset \and M.~Tagger}

\offprints{F.~Masset}
\thesaurus{307, 21--32 (1996)} 

\institute{CEA, DSM, DAPNIA, Service d'Astrophysique, CE-Saclay, 91191 
Gif-sur-Yvette Cedex, France} 

\date{Received 6 October 1994 / Accepted 21 July 1995}
\maketitle

\begin{abstract}
We show that the propagation of warps in gaseous disks can be strongly 
affected by compressional effects, when the thickness of the disk is taken 
into account. The physical reason is that, in realistic 
\new{self-gravitating} disks, the sound 
time through the disk is comparable with the rotation time; thus the 
vertical hydrostatic equilibrium cannot be maintained adiabatically 
\new{as the wave propagates} (an 
implicit hypothesis in the thin-disk approximation) and the disk cannot 
move up and down solidly to follow the warp perturbation. There results, 
\new{together with the main vertical motion}, a
strong horizontal one which significantly modifies  
the dispersion relation. 

We then turn to the case of a disk composed of two fluids with 
different temperatures: this can correspond either to the combined 
motions of the gas and the stars in a galactic disk, or to their coupling 
with a flattened massive halo (assuming that, as for spiral waves, stars 
are conveniently represented by a fluid if one stays away from Lindblad 
resonances). We find, in addition to the usual warps, a short-wavelength 
wave which might explain the ``corrugations'' observed in many galactic disks.

\no{Finally, as a side result of this analysis, we discuss 
a possible weak amplification of $m> 1$ warps.}


\end{abstract}

\section{Introduction}
Warps are bending waves of disk galaxies, observed most often in the outer 
parts of edge-on spiral galaxies in the 21 cm line of neutral hydrogen. 
Their physics is very similar to that of spiral density waves, \new{from 
which they are distinguished by their opposite parity:} they 
involve vertical rather than horizontal motions (\neuf{though we shall see
that horizontal motions in warps can be important as well})
and antisymmetric (in the 
vertical coordinate) rather than symmetric perturbed potentials. Their 
propagation has been described in the limit of very thin disks, and found 
to explain the main observational properties. Warps have also been 
considered as a possible source of angular momentum transfer in accretion 
disks. On the other hand, contrary to spirals, warps are not unstable, \ie 
they cannot form spontaneously in an isolated disk.
Since warps propagate radially in the disk, a continuous 
excitation mechanism must be found to explain their frequent occurrence in 
galactic disks; various mechanisms have been considered (see~\eg Binney, 1992): \begin{itemize}
\item{tidal excitation of warps (Hunter and Toomre 1969) by a companion 
galaxy; this cannot explain
the observation of isolated warped galaxies, and seems too weak to explain 
the observed amplitude of galactic warps. This has recently raised a 
strong interest in the context of accretion disks (Terquem 1993).} 
\item{The bending of the trivial tilt mode by an elongated halo whose axis 
is misaligned with the axis of the galactic disk (Sparke 1984, see also 
Sparke and Casertano 1988). Although this hypothesis can describe very 
accurately the observed shape of prototype warps, it raises the problem of 
the origin and maintenance of such a misalignment. Recently Dubinski and 
Kuijken (1995) found that the axis of the halo and the disk should rapidly 
realign.}
\end{itemize}

This is the first in a series of papers where we will present a non-linear 
excitation mechanism for warps from spiral waves in disk
galaxies. As a preliminary to this non-linear work we will concentrate 
here on the linear properties of warps; we find important differences with 
the classical theory, when we take into account the finite thickness of 
the disk. The reason is that the thin-disk approximation used in the 
classical theory relies on an underlying hypothesis, that the whole gas 
column at a given location in the disk can move up or down solidly, 
without changing its vertical profile; this would require that the 
hydrostatic vertical equilibrium can evolve adiabatically, \ie that the 
sound time through the disk, $\tau_s=H/a$ where $H$ is the disk scale 
height and $a$ the sound speed, be shorter than the typical time scale of 
the warp, which is of the order of the rotation time $\tau_R=\Omega^{-1}$. On the 
other hand the vertical hydrostatic equilibrium implies $\tau_s\sim \tau_R$
if the vertical gravity is dominated by a central object, or -a more 
appropriate condition for the outer galactic disk- if it is dominated by 
the self-gravity of the gas or the stars, provided that the Toomre 
parameter $Q$ is of the order of one. 

\no{Finite thickness and compressibility effects have been intensively studied
for spiral waves (Shu~1968, Vandervoort~1970, Romeo~1992,~1994). As
an important result they found in particular that the critical Toomre
parameter $Q_{crit}$ for disk stability is lower than unity, as predicted
by the thin disk analysis, at most by a factor about~2.}

Thus finite thickness 
and compressibility effects must also be taken into account for a realistic 
description of the warps. We find that they can substantially modify the 
dispersion relation and that they introduce strong horizontal motions, 
comparable with the vertical one. 

\no{Our method is similar to that of Shu (1968) for spiral
waves, apart from modifications
for taking into account the opposite parity of warps.}

\new{On the other hand galactic disks are characterized by distinct  
populations, with different temperatures and scale heights, contributing 
separately to the vertical potential well: one has stars and gas, 
and also the halo whose contribution is believed to dominate the gravitational 
potential. In a first approach of the complexities introduced by these 
distinct populations, we present an analysis of warps in a disk composed 
of two fluids with distinct temperatures: assuming that, as for spiral 
waves, stars can be conveniently represented as a fluid if one stays away 
from the Lindblad resonances, this allows us to discuss the case of warps 
in a disk of gas and stars, or in a disk embedded in a flattened halo. We 
find that, in addition to the usual warp wave, a new one exists which 
might correspond to the ``corrugations'' observed in many galactic disks
\no{(Quiroga \etal~1977, Florido \etal~1991)}.}

Finally we will also present an extension of this analysis
to a weak amplification mechanism for $m> 1$ warps. This 
is of very limited interest for galactic disks, where the amplification time would at 
best be a sizable fraction of the Hubble time, but might apply to 
accretion disks.

\section{Notations and formalism}

The dispersion relation of warps in an infinitely thin disk is (Hunter 
and Toomre 1969, Binney and Tremaine 1987) 
\begin{equation}
\label{eqn:RD}
\tomega^2 = \mu^2 + 2\pi G\Sigma q
\end{equation}
where $\tomega=\omega-m\Omega(r)$ is the warp frequency in the frame of 
the matter, $m$ is the azimuthal \no{pattern number}, $\mu$ is the  vertical 
oscillation frequency of the disk, $\Sigma$ is the mass per surface unit 
of the disk, and $q$ the modulus of the warp's horizontal wavevector. 

We want to derive the dispersion relation for a moderately thick disk, 
\new{\ie a disk which is geometrically thin (\(H/r\ll 1\)), where \(H\) 
is the disk thickness), but where the ratio \(\tau_S/\tau_r\) can be of 
the order of one}. This will be done in a manner similar to the usual thin 
disk derivation, 
but we now have to consider the three projections (radial, azimuthal and 
vertical) of the Euler equation, and the solution of the Poisson equation 
becomes more complex.

\subsection{Notations}

Let us first introduce some notations.
We use the shearing
sheet model to describe differential rotation (Goldreich and Lynden-Bell 
1965). In this model one considers an annulus around corotation as a 
cartesian slab with $x=r-r_0$ (where $r_0$ is the corotation radius where 
$\tomega=0$) and $y=r\vartheta$. The radial variations of all equilibrium 
quantities are neglected, except the rotation speed $V_0(x)\vec{e_y}$ 
which varies linearly, $V_0 = r_0 \Omega(r_0) + 2Ax$ where $A = 1/2 
r\partial \Omega / \partial r$ is Oort's first constant. We
also restrict ourselves to the case of isothermal equilibrium and 
perturbations, with sound speed $a$.

In what follows we denote by $U$, $V$ and $W$ the components of the perturbed 
velocity, $\rho$ the perturbed density and $\phi$ the perturbed potential. 
We call $k_x$ and $k_y$ the projections of the wavevector of the warp we 
study, $q=(k_x^2+k_y^2)^{1/2}$ its modulus, $\tomega$ its frequency.
\no{The connection between the cylindrical geometry and the shearing sheet
is given by $k_y=m/r_0$.} 
Unperturbed quantities are denoted by the same symbols with a subscript 0. 

\subsection{The formalism}

The dispersion relation is derived in a WKB approximation, whose validity 
condition is the ``tightly wound'' approximation, $k_x\gg k_y$. In section 
5 we will give a numerical solution, independent of this approximation.

We limit ourselves to a linear analysis of the propagation of warp waves, 
i.e. perturbed quantities are infinitesimal.
With these restrictions, it is possible to characterize a warp wave by 
parity considerations, since
the parity of the equilibrium and of the perturbation equations allows one 
to separate symmetric and antisymmetric solutions. In the former, 
corresponding to spiral density waves, all perturbed quantities are even 
in $z$, except $W$ which is odd. The opposite holds for antisymmetric 
solutions, corresponding to warps.

\subsection {The basic equations}
We start from the continuity
equation and the horizontal projections of the Euler equation: 

\begin{equation}
\label{eqn:conti0}
-i\tomega \rho + \rho_0(ik_xU+ik_yV)+\frac{\partial}{\partial z}(\rho_0 
W) = 0
\end{equation}

\begin{equation}
\label{eqn:eulerX}
-i\tomega U - 2\Omega V=-ik_x\biggl(\phi+a^2\frac{\rho }{\rho_0}\biggr)
\end{equation}

\begin{equation}
\label{eqn:eulerY}
-i\tomega V + \frac{\kappa^2}{2\Omega} U=-ik_y\biggl(\phi+a^2\frac{\rho 
}{\rho_0}\biggr)
\end{equation}

where $\kappa=(4\Omega^2+4\Omega A)^{1/2}$ is the epicyclic frequency. 
Using the WKB assumption ($k_x \gg k_y$) we get:

\[k_xU+k_yV = 
q^2\frac{\tomega}{\tomega^2-\kappa^2}\biggl(\phi+a^2\frac{\rho} 
{\rho_0}\biggr)\]

Thus the continuity equation~(\ref{eqn:conti0}) can be written as: 

\begin{equation}
\label{eqn:da}
\frac{\partial\alpha}{\partial z} = \biggl[-\tomega s + \frac{q^2\tomega} 
{\tomega^2-\kappa^2}(\phi + a^2 s)\biggr]\rho_0(z) \end{equation}

where

\[\alpha = i\rho_0(z)W\]
\[s=\frac{\rho}{\rho_0}\]

Our basic set of equations also includes the vertical projection of the 
Euler equation: 

\begin{equation}
\label{eqn:ds}
\frac{\partial s}{\partial z} = \frac{1}{a^2}\biggl[\frac{\tomega\alpha} 
{\rho_0} - \frac{\partial\phi}{\partial z}\biggr] \end{equation}

and the Poisson equation:

\begin{equation}
\label{eqn:d2phi}
\frac{\partial^2\phi}{\partial z^2} = 4\pi G\rho_0s + q^2 \phi 
\end{equation}

In equation (\ref{eqn:da}) the second term in the bracket represents the horizontal 
part of the divergence of the velocity field, \ie the compressional 
contribution which is the main new physical effect introduced in this 
paper. For $\tomega\sim \Omega\sim\kappa$, and unless $\tomega$ is very 
close to $\kappa$ (\ie at the Lindblad resonances), this term is of order 
of $q^2a^2/\kappa^2$; this is of the order of $q^2H^2$ compared to the 
first term, if the disk is at hydrostatic
equilibrium, with a gravity either dominated by a central object or (if 
Toomre's parameter $Q=\kappa a/\pi G \Sigma\sim 1$) by the local gravity 
of the gas, so that $H\sim a/\kappa$. We will return to this point in the 
following sub-section. 
Restricting ourselves to wavelengths larger than the disk thickness, we 
will use $qH$ as the expansion parameter in the following analysis.

\new{On the other hand, we will ignore here the case, which might apply to 
galaxies, where gravity is dominated by a massive halo, so that $H\ll 
a/\kappa$. If the halo is passive, 
\ie does not participate in the motion, the disk is indeed thin in the 
sense that $\tau_S\ll\tau_R$.
On the other hand, if the halo is flattened it presumably 
participates in the differential rotation and can also be involved 
dynamically in the 
warp. Its effect will be discussed below, in the section devoted to warps 
in a two-component disk. Even in the case of a passive halo, we will find 
new effects associated with corrugations of the disk.}

\subsection{\nouveau{Consistent model of geometrically thin disk}}
{\new{In the following
we will make extensive use of consistent models of disk vertical density 
and potential profiles. We build them as coupled solutions of the 
hydrostatic equilibrium and the Poisson 
equations, in the hypothesis of a geometrically thin disk ($H/r\ll 1$).

More precisely, let us construct 
{\em ab nihilo} such a consistent disk. In a first guess, we choose the  
density profile $\rho_0(z)$. The vertical hydrostatic equilibrium gives 
the gravitational potential by:}}

\[\rho_0\frac{\partial 
\phi_0}{\partial z}=-a^2\frac{\partial}{\partial z}\rho_0(z)\]
{\new {On the other hand the Poisson equation gives:}} 
\[{1\over r}{{\partial }\over{\partial r}}\biggl(r\frac{\partial} 
{\partial r}\biggr)\phi_0=4\pi G\rho_0 - \frac{\partial^2\phi_0} {\partial 
z^2}\]

{\new {where we have written what we know (\ie what is imposed by our 
choice of $\rho_0$) in the R.H.S. Thus the Poisson equation 
gives the behavior of the L.H.S., which is the {\em radial} part of the 
laplacian, hereafter denoted by $\Delta_r$:}}

\[\Delta_r={1\over r}{{\partial }\over{\partial r}}\biggl(r\frac{\partial} 
{\partial r}\biggr)\]

{\new {It is an easy matter to check 
that this radial laplacian is equal to:}}

\begin{equation}
\label{eqn:muHT}
\Delta_r\phi_0 = \kappa^2-2\Omega^2 \equiv - \mu^2
\end{equation}

Note that $\mu^2$, that we have called the vertical oscillations frequency,
does not involve any term in $4\pi G\rho_m$. Indeed 
$4\pi G\rho_m$ appears when considering the vertical oscillations of a test
particle in the rest potential of the disk, whereas we are concerned here
with global oscillations which involve vertical motion of the potential
well itself.
{\new{(see Hunter and Toomre, 1969, for a derivation of equation [\ref{eqn:muHT}]). Since the density profile is 
arbitrary, so is $\mu^2$. But in  a {\em geometrically thin} disk $\Omega$ 
and $\kappa$ 
must be independent of
$z$. Thus  $\Delta_r\phi_0$ must also be constant with respect to $z$. 
We can thus summarize the definition of a consistent disk equilibrium as a 
disk whose density profile obeys the equation}} 

\begin{equation}
\label{eqn:lapphi}a^2\frac{\partial}{\partial 
z}\biggl(\frac{1}{\rho_0}\frac{\partial \rho_0} {\partial 
z}\biggr)=\Delta_r\phi_0-4\pi G\rho_0 \end{equation}
{\new{Where both $a$ and $\Delta_r\phi_0$ are independent of $z$. 

By choosing them we can construct a continuous palette of 
vertical equilibria, from a keplerian disk (where the vertical gravity is 
dominated by the mass of a central object, \ie where the first term is 
dominant in the R.H.S. of equation [\ref{eqn:lapphi}]) to self-gravitating 
ones, where gravity is dominated by the local distribution of matter 
(\ie where the second term is dominant in equation [\ref{eqn:lapphi}]). 
One easily obtains in the first case a gaussian profile 
$\rho_0(z) = \rho_m\exp(-z^2/H^2)$ where $\mu^2=-\Delta_r\phi_0 = 
2a^2/H^2$ (and in that case, $\mu$ and $\kappa$ are equal to $\Omega$). 
The second case straightforwardly leads to: 
$\rho_0(z)=\rho_m/\cosh^2(z/H)$ where $H=a/\sqrt{2\pi G\rho_m}$. %
Between these extremes, one can consider any mixture of local and 
global gravity.}} 

{\nouveau {Since the construction of a consistent disk implies the choice
of two independent parameters (the sound speed $a$ and $\mu^2$), the set of
consistent disk equilibria is a two-dimensional manifold where any
independent couple of parameters can be chosen to ``label'' a particular 
choice; We will use below $\mu/\kappa$ and the Toomre parameter 
$Q=a\kappa/\pi G\Sigma$.}}

{\new {Then obviously the notion of consistent disk equilibrium does not imply 
any constraint on the Toomre parameter $Q$, and on the ratio 
$\tau_S/\tau_R$ discussed in the introduction.}}

{\nouveau {However, in order to build realistic disk equilibria, one 
cannot choose $\mu/\kappa$ and $Q$ arbitrarily: indeed if $\mu/\kappa$ is 
close to one, meaning that the disk
is nearly keplerian and  dominated by the gravity of the central
regions, the $Q$ parameter must be large enough to ensure that
the local gravity is low enough.
On the other hand, if $\mu/\kappa$ is not close to one,
$Q$ must not be too large, so as to enable the
local gravity to play a role in the equilibrium of the disk. Let us also 
recall, as already mentioned, that we will not discuss here the 
contribution of a halo to the vertical gravity and thus to the frequency 
$\mu$. This discussion is deferred to the consideration of two-component 
disks in section 4.

In the following we study the dispersion relation of warps for various
type of disks, from self-gravitating to keplerian. We label 
them by the quantity $\mu/\kappa$. Of course we have checked
that the Toomre parameter of these disks is realistic in the sense
that it is low for self-gravitating disks and very large in the case of the
keplerian disk.}}

\section{The dispersion relation of warps} 

We write the dispersion relation by solving the linear system (5--7) for 
$\alpha$, $s$ and $\phi$, subject to the appropriate boundary conditions. 
We have four such conditions, two at the mid plane ensuring the parity of 
the solution:

\[s=0\]
\[\phi=0\]

and two at infinite $z$:

\[\alpha = 0\]
\[\frac{\partial \phi}{\partial z} = -q\phi\] 

which ensure that the mathematical solution which satisfies the 
warp parity is also a physical one, with an exponential decrease of the 
perturbed potential, and a vanishing momentum flux due to the rarefaction 
of matter. A fifth condition corresponds to the amplitude of 
the solution (arbitrary since we consider linearized equations). Since we 
solve a fourth-order system, these conditions can 
be fulfilled only if the parameters $q$ and $\tomega$ obey a condition, 
which is the dispersion relation. \toutnouveau{This dispersion relation can also be
seen as a compatibility condition of an overdetermined differential
system which has both Dirichlet's and Neumann's boundary conditions,
as explained in Bertin and Casertano (1982), who investigated the dispersion
relation of bending waves in incompressible thick disks with constant 
thickness.}

\subsection{Numerical determination of the dispersion relation} We 
determine numerically the dispersion relation by solving the system (5--7) 
and looking for the values of $q$ and $\tomega$ which allow the solution 
to obey the boundary conditions. Our procedure is the following: for a 
given value of $q$ and choosing an approximate value of $\tomega$ we 
integrate the system (5--7) from large $z$ to zero. Obeying the boundary 
conditions at infinity means that we start with $\alpha=0$ and $\partial 
\phi/\partial
z=-q\phi$ (note that an error in the latter condition will give a 
projection on the solution that diverges at large $z$, so that as we 
integrate toward decreasing $z$ its effect will be exponentially small); 
we choose $s=1$ and we still have one free parameter, the 
value of $\phi$. We do this twice, starting with different values of 
$\phi$, and resulting in two different sets of values of $s$ and $\phi$ at 
$z=0$. We can then find a linear combination of these two solutions to 
form a third one, which still obeys the conditions at infinity and now 
also has $s(0)=0$. Then a Newton method allows us to vary the value of 
$\tomega$, for a given $q$, until the last condition,
$\phi(0)=0$ is also satisfied. This results in $\tomega(q)$, the 
dispersion relation.

\subsection{Analytical derivation of the dispersion relation} 
\no{We remind that equation~(\ref{eqn:RD}) has been obtained for
an infinitely thin sheet of matter involving only
vertical motion. Thus this dispersion relation does not take
into account:
\begin{itemize}
\item{{\bf{compressional effects}} due to the finite sound speed (the finite 
ratio $\tau_S/\tau_R$).
We will emphasize this 
point further below, when we derive the analytical dispersion relation for 
a moderately thick disk;} 
\item{{\bf{horizontal motions}};}

\item{{\bf{geometric effects}}, often approximated by ``softened gravity'' 
models for spiral density waves. Analytical calculations for 
an incompressible disk in which horizontal motion is suppressed show - for 
any density profile - that $\tomega$ tends to an asymptotic value 
$\tomega_{\infty}$ when $qH$ becomes large which is obviously not
the case with equation~(\ref{eqn:RD}).}

\end{itemize} 

We present a method which allows us to derive analytically 
an approximate dispersion relation, taking into account 
the compressional effects and the finite thickness
for small but finite values of $qH$.} It must be emphasized that 
the thin-disk dispersion relation gives

\[q\sim \Omega^2/2\pi G \Sigma\sim \Omega/a\] 

for $\tomega\sim \Omega$, and a disk which has either a low self-gravity, 
or a strong one with the Toomre parameter $Q=\kappa a/\pi G \Sigma \sim 1$ 
(the latter case being more relevant for the outer regions of disk 
galaxies). Thus $qH\sim \Omega H/a=\tau_s/\tau_R$, the ratio of the sound 
time through the disk to the dynamical time: the thin disk approximation 
($H\rightarrow 0$) involves an assumption that the sound time through the 
disk is small, so that the vertical hydrostatic equilibrium can be 
maintained throughout the evolution of the waves of interest and 
compressional effects can be neglected. On the 
contrary the new effects we find here, which are associated with the 
finite value of $qH$, result from the fact 
that this equilibrium cannot be maintained perfectly. 

Our method, which aims at
eliminating the $z$ dependence
of the variables, consists in making an integral over $z$ which represents 
the amplitude of the warp (the mean displacement from the galactic plane), 
and transforming it by using equations (\ref{eqn:da}--\ref{eqn:d2phi}). 
The main difficulty comes from the potential. Here we express it by using 
Green's functions, which allows us to separate the vertical structure of 
the solution without having to average over $z$, as is classically done. 
We show that it is possible to modify the dispersion relation by a 
second order term in $qH$. This term, similar to the pressure ($a^2q^2$) 
term of the 
dispersion relation of spiral modes, contains the contribution of the 
compressional effects, but does not contain the modification of the gravity
force, which is more difficult to evaluate, and would require the 
analytical knowledge of the eigenvectors.

\no{The derivation can be found in appendix, and it leads to:}

\[\tomega^2 = -\Delta_r\phi_0 + 2\pi G \Sigma q + 
\frac{\tomega^2}{\tomega^2-\kappa^2}a^2q^2 + O_{grav}[(qH)^2].\]

\no{where $O_{grav}[(qH)^2]$ represents the (unknown) contribution
of self-gravity at orders~2 and higher in~$qH$.}

Using equation~(\ref{eqn:muHT}), and neglecting 
the second-order gravitational contribution, we finally obtain: 

\begin{equation}
\label{eqn:RDM}
\tomega^2 = \mu^2 + 2\pi G \Sigma q + 
\frac{\tomega^2}{\tomega^2-\kappa^2}a^2q^2. 
\end{equation}

This dispersion relation is identical to the thin-disk 
one~(\ref{eqn:RD}), except for the 
second-order compressional term. This term comes from the horizontal 
motions driven by the horizontal gradients of the perturbed pressure and 
potential, and diverges at the Lindblad resonance where epicyclic motion 
is resonantly excited by the pressure force. However this divergence must 
be taken with caution: we have thus far discussed only the case of a 
gaseous disk, through the use of hydrodynamical equations. It is well 
known that the results for a collisionless stellar disk are usually very 
similar, except in the vicinity of the Lindblad resonances where a major 
difference occurs between the two cases: the stars resonate with the wave 
and can exchange energy with it through Landau damping, whereas the gas 
does not. Here the divergence of this term does not mean a resonance, but 
rather the fact that our expansion fails in this region. Still, we note 
that this allows us to get close enough to the Lindblad resonance that our term starts playing an important role, for stars as well as for the gas.

We also wish to note that a similar 
contribution to the dispersion relation had been found by Nelson (1976 and 1981) in 
the limit of weak shear, and more recently by Papaloizou and Lin (1995) in 
a more general study in cylindrical geometry. 

\subsection{Physical interpretation of the additional term} 
\begin{figure}
\psfig{file=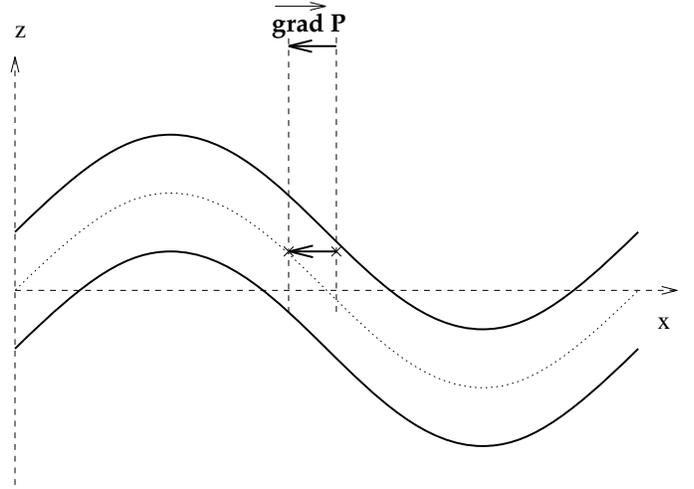,width=\columnwidth} 
\caption{\label{fig:interpret} \scriptsize In this figure we have assumed 
a purely solid vertical motion. Hence the density is conserved and 
advected by the vertical motion, and, since pressure is linked to 
density by: $P=a^2\rho$, horizontal gradients of density and pressure 
arise in the disk, due to the vertical stratification of the equilibrium 
configuration. These gradients are proportional to $q$, the horizontal 
wavenumber, and to $H^{-1}$, the vertical gradient.} \end{figure}

\no{The 
additional term in the dispersion relation is linked to 
the divergence of the horizontal perturbed velocity
(as can be seen in the derivation given in appendix, from
equation~[\ref{eqn:S}])}. Hence this additional 
term is due to the presence of horizontal, compressional motions associated
with the warp. In order to 
understand how these horizontal motions arise, let us first consider a 
warp in which motion is purely vertical and solid. This situation is 
presented in figure~\ref{fig:interpret}. 

\begin{figure}
\psfig{file=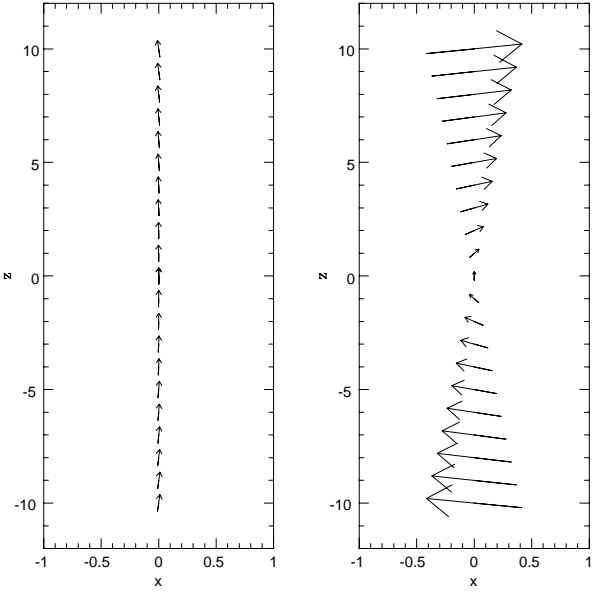,width=\columnwidth} \caption{\scriptsize 
\label{fig:vfield}
We show (left) the velocity field $(U,W)$ of a warp far from the 
Lindblad resonance, and (right) the same field close to  
the Lindblad resonance. If we define $R$ as the ratio of the $2\pi G\Sigma 
q$ term in the dispersion relation over the additional 
$a^2q^2\tomega^2/(\tomega^2-\kappa^2)$ term, we have $R=41$ 
in the first case, and $R=0.53$ in the second case. These fields have been 
obtained numerically 
by the method described in the text. The disk scale height $H$ is about $2$ in 
these plots.}
\end{figure}

{\new{It shows how 
horizontal pressure gradients appear in the system, which in turn 
tend to move matter horizontally. The amplitude of horizontal motions is 
greater and greater as the warp frequency in the frame of matter 
($\tomega$) approaches the frequency at which matter spontaneously moves 
horizontally, \ie the epicyclic frequency. This can be the beginning of a 
justification of the resonant denominator in the 
additional term to the dispersion relation. In order to illustrate the 
behavior of matter far and close to the Lindblad resonances, we have 
plotted in figure \ref{fig:vfield} the velocity fields of matter in these 
two cases. We see that far from the Lindblad resonance, the 
matter moves essentially vertically with the motion described by the 
dispersion 
relation without our additional term; close to the Lindblad 
resonance, on the other hand, the motion tends to be practically 
horizontal as soon as we 
leave the median plane. In this case, the motion might be described by 
two slices moving in opposite directions, giving a strong vertical 
shear.}} 

{\new{In a forthcoming paper discussing the possible excitation of warps by 
spiral waves, we will fully analyze the energetics of warps, and in 
particular we will discuss the fraction of the total energy of the warp 
that can be stored in horizontal motions. 

It is noteworthy that our additional term is linked to the finite thickness of 
the disk, since a fundamental hypothesis is that the disk equilibrium is {\em 
consistent}, and since the sound speed which appears in that additional 
term is proportional to the thickness of the disk.}}

\subsection{Comparison between analytical derivation and numerical 
results} 
\begin{figure}
\psfig{file=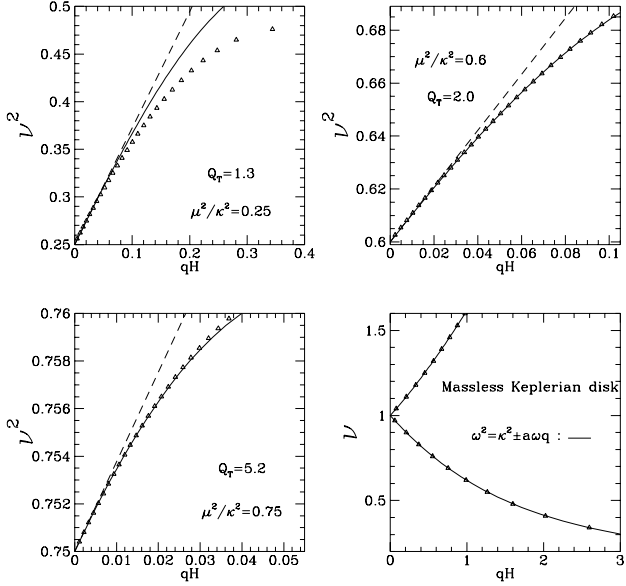,width=\columnwidth}
\caption
{\label{fig:r2}
\scriptsize Comparison between the analytical and numerical dispersion 
relations. The quantity $\nu$ represents $\tomega/\kappa$. The 
triangles represent the
results of numerical integration, the dashed line the infinitely thin disk 
dispersion relation, and the solid line the thick disk dispersion 
relation. 
{\new{For each curve we have chosen a 
range of $qH$ such that the diverging trend 
between the numerical dispersion relation and the infinitely thin disk 
dispersion relation appears at about the first quarter of the $qH$ range. 
The ordinate in the bottom right diagram is $\nu$ 
rather than $\nu^2$}}. See text for explanations.} 
\end{figure}

\no{We present the comparison between the numerical dispersion
relation and the dispersion relation given by
equation~(\ref{eqn:RDM}) in figure~\ref{fig:r2}. 
We obtain a very good agreement between the numerical and 
analytical solutions, which appears to be far better than
the infinitely thin disk dispersion relation~(\ref{eqn:RD}),
except in the top-left diagram, where the disk is 
strongly self-gravitating, so that the gravitational part of the 
second-order term, which we did not evaluate, cannot be neglected.}

We have also emphasized in this figure the case of a massless keplerian 
disk which is for us, according to our definition of consistent disk, a 
{\it massless} disk ($\Sigma = 0$), for which $\Omega \propto r^{-3/2}$. 
For such a disk, the general dispersion relation becomes: 

\[\tomega^2=\kappa^2\pm a\tomega q\]

This dispersion relation is exact , since the unknown correction term in 
our derivation was due to self-gravity, which we do not need to take into 
account for a keplerian disk. Thus we can see that the warp of a keplerian 
disk precesses, a result which was not given by the infinitely thin disk 
relation which in that case becomes: \[\tomega^2=\kappa^2=\Omega_K^2\]

The precession frequency, expanded to lowest order in $qH$, is thus of the 
order of \[\Omega_K {H\over R}\]
(for a wavelength of the warp of the order of the size of the disk). This 
can have important consequences in the context of protoplanetary disks. 
The role of warps has recently raised a strong interest, based in part on 
the fact that in keplerian disks the warp is a neutral tilt mode which has 
a vanishing energy and can thus very easily be excited. We expect that our 
effect should involve a finite warp energy, and thus make excitation 
mechanisms less efficient.

\no{A similar resolution for the spiral waves dispersion relation 
(where we omit again the second and higher orders terms in $qH$
for the self-gravity) leads
to the infinitely thin disk dispersion relation;

\begin{equation}
\label{eqn:RDS}
\tomega^2=\kappa^2-2\pi G\Sigma q+a^2q^2
\end{equation}

From this result we can say that equation~(\ref{eqn:RDS}) and
equation~(\ref{eqn:RDM}) describe respectively the propagation
of spiral waves and of bending waves with the same accuracy,
\ie with the same underlying hypothesis.}

\section{Two-fluid dispersion relation}

In this section we derive the dispersion relation of the bending waves in 
a two-fluid
system. Both fluids have different sound speeds and
hence scale heights, and they have independent surface densities. Our 
primary goal is 
to discuss warps in a disk composed of stars and gas 
(assuming that the stars are correctly represented as a warm fluid), or 
to a disk embedded in a flattened halo which can participate dynamically 
in the warp. In the following, for simplicity, we refer to the two fluids 
as gas and stars.

We will find in particular that, in a star-gas disk, we find solutions 
closely resembling the corrugations, \ie short wavelength bending waves 
of the gas layer observed in
certain edge on spiral galaxies and in the Galaxy. 

We keep the same notations as for the mono-fluid case, with an index
$*$ or $g$ applying respectively to the warmer (stars) and the cooler 
(gas) species.

We still need the assumption of consistent vertical equilibrium already used in
our mono-fluid derivation, \ie that both fluids obey the Poisson
equation and the hydrostatic equilibrium, while the 
gravitational potential 
is such that $\Delta_r\phi_0$ is constant through the disk
thickness.

\no{With a method strictly similar to that exposed in appendix for the
monofluid case, we can write down equations for $N_*$ and $N_g$,
which constitute a linear homogeneous system.
This
system will have a non-trivial solution only if it has a
vanishing determinant.

Hence the dispersion relation of warps in a two-fluid system is:

\begin{equation}
\label{eqn:rd2f}
{\cal D}_g{\cal D}_*-\biggl(\frac{2\pi G\Sigma_gq}{\tomega^2}
-\frac{\nu_{*}^2}{\tomega^2}\biggr)\biggl(\frac{2\pi G\Sigma_*q}{\tomega^2}
-\frac{\nu_{g}^2}{\tomega^2}\biggr) = 0
\end{equation}

where we have defined:

\[{\cal D}_i=\frac{1}{S_i}
-\frac{\mu^2+\nu_{i}^2+2\pi G\Sigma_iq}{\tomega^2}
\mbox{  where $i=*$ or $g$}\]

and:  

\[\nu_{i} = 4\pi G\frac{\int_0^{+\infty}\rho_j\alpha_i}{N_i}
\mbox{   where $j \ne i$}\]

Physically, ${\cal D}_i=0$ would be the dispersion relation of bending waves if
there was only the species~$i$. 
The role of $\nu_g$ becomes clear if one goes to the 
limit where the gas disk is much thinner than the stellar one: in that 
case one has $\nu_g = 4\pi G\rho_{m*}$, where $\rho_{m*}$ is the stellar
density at the disk mid-plane; $\nu_g$ then appears as the vertical 
oscillation frequency of the gas in the stellar potential.}

\neuf{It must be noted that realistic stellar disks can have
a vertical velocity dispersion very different from the radial
one. This might affect the present results by numerical factors
which could be important in detailed comparisons with observations.
This will be considered in future work.}

For a given $\tomega$, equation~(\ref{eqn:rd2f}) 
is of order $4$ in $qH$ (since the $1/S_i$
hidden in the ${\cal D}_i$ 
are of order $2$ in $qH$). In general its roots cannot be easily 
expressed, but we find it interesting to first analyze them in the 
trivial case $a_g=a_*$, before turning to numerical solution in the 
general case.

\subsection{Roots identification of the two-fluids dispersion
relation}
\subsubsection{Peculiar case $a_g=a_*$.}
In this case we have $S_g=S_*\equiv S$, and we are dealing with two physically 
indistinguishable fluids.
Now we can factorize the dispersion relation (\ref{eqn:rd2f}).
We obtain after some straightforward
transformations:

\[\biggl(\frac{1}{S}-\frac{\mu^2+2\pi G\Sigma_tq}{\tomega^2}\biggr)
\biggl(\frac{1}{S}-\frac{\mu^2+\nu_g^2+\nu_*^2}{\tomega^2}\biggr)=0\]

where $\Sigma_t=\Sigma_*+\Sigma_g$.
The eigenvector associated with the second factor can be easily found
by summing equations relative to $N_*$ and $N_g$, giving:
\[\biggl(\frac{1}{S}-\frac{\mu^2+2\pi G\Sigma_tq}{\tomega^2}\biggr)(N_g+N_*)=0\]
Thus when the second factor vanishes this gives:
$N_g+N_*=0$.
This condition consists in splitting the gas in two parts 
that move in opposite directions, so that the ``total'' warp (the average 
displacement) vanishes. In 
fact we note that the second factor is of the form  $aq^2+b$, so 
that it gives only one positive root for $q$.  We will call this the 
``hidden'' mode, even when we 
relax the condition $a_g=a_*$.

In the same manner, it is an easy matter to check that the 
eigenvector
of the first factor is $\Sigma_*N_g-\Sigma_gN_*=0$. This means that both 
fluids have the same behavior
(they have in particular the same elongation $Z$), and we recover the 
one-fluid dispersion relation.

Finally we have three modes: the two ``classical'' modes of the one
fluid disk, given by the second order dispersion relation
of a one-fluid warp, and the ``hidden mode'' associated to
the second factor.

\subsubsection{General case}
When $a_g$ and $a_*$ are different, the simple factorization
seen above is no more possible and we turn to numerical solution.

We have adopted the following values: 
\[\mu^2/\kappa^2 = 0.1 \]
(corresponding to a nearly flat rotation curve)
\[\nu_g^2/\kappa^2 = 10 \]
\[\tomega/\kappa = 0.75 \]

(so that we are between the warp's Lindblad resonance (inner or outer) and the forbidden
band around corotation where the warps do not propagate)

\[\Sigma_g/\Sigma_* = 0.1\]

\begin{figure}
\psfig{file=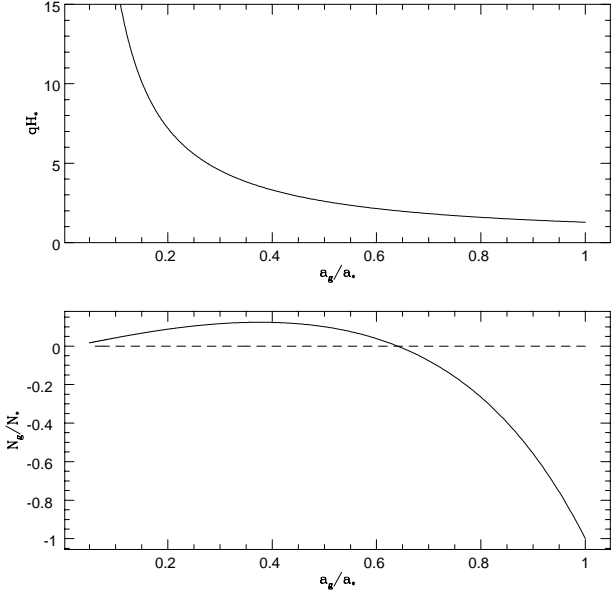,width=\columnwidth}
\caption
{\label{fig:qp} 
\scriptsize The dimensionless wavenumber of the hidden
mode for values of parameters
typical of our Galaxy's, and the ratio $N_g/N_*$, which
is ten times ($\Sigma_*/\Sigma_g$) the ratio of deviations
of the stars and gas.
The ratio $N_*/N_g$ vanishes at $a_g/a_*=0.6524$ (i.e. at $qH_*=2$),
which corresponds
to a regime where the stars are strictly motionless.}
\end{figure}

We also need to choose the value of $\nu_*$. It is
an easy matter to check that, since we limit ourselves to the
(realistic) case $a_g \ll a_*$:

$\nu_g\simeq 4\pi G\Sigma_*/H_*$ and
$\nu_* \simeq 4\pi G\Sigma_g/H_*$ hence $\nu_*
\simeq \nu_g\Sigma_g/\Sigma_*$.

We will not discuss the standard warp mode, which behaves in a manner 
very similar to the one-fluid case. The stars and gas motions are nearly 
identical, though since $a_g\neq a_*$ small differences occur. We will 
rather discuss the behavior of the ``hidden'' mode, and tentatively 
identify it with the corrugations observed in many galaxies.

The results are plotted on figure \ref{fig:qp}. The top plot shows $qH_*$, 
while the bottom one shows the ratio $N_*/N_g$, 
as a function of
$a_g/a_*$ for the hidden mode.
When the sound speeds are identical we check the result of the previous 
subsection, that the two fluids
move in opposite directions so as to achieve a null global vertical
displacement. When the gas sound speed
is decreased, the ratio $|N_*/N_g|$ decreases, \ie 
the stars are more and more motionless. Thus the mode under
study is essentially gaseous. The stars become passive, though they still 
act through their unperturbed potential to change the frequency of 
vertical oscillations of the gas disk. One may note that for 
$a_g/a_*<.65$ the ratio $N_*/N_g$ changes sign, \ie the stars now move in 
the same direction as the gas.

\begin{figure}
\psfig{file=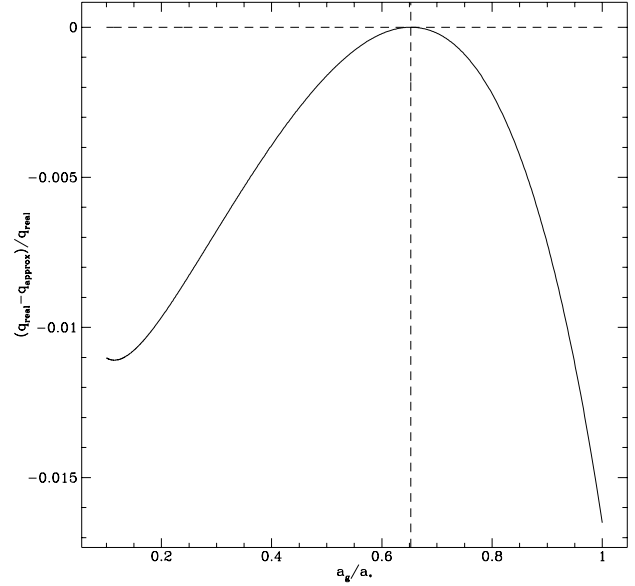,width=\columnwidth}
\caption
{\label{fig:err} 
\scriptsize Plot of the relative error between the exact numerical solution and the approximate
dispersion relation. We see that the error remains weak, and vanishes when
$a_g/a_*=0.6524$, which is logical since at this point the stars are 
strictly motionless.}
\end{figure}

We have monitored the relative error obtained with the computation
of $q$ by the approximate dispersion relation:

\[\tomega^2 = \mu^2+\nu_g^2+2\pi G\Sigma_gq+\frac{\tomega^2}
{\tomega^2-\kappa^2}a_g^2q^2\]

\ie the one-fluid dispersion relation where we have added
the vertical frequency $\nu_g$ due to the rest potential of the stars.

This error is plotted on figure \ref{fig:err}. We see that it is always
very weak, so that the approximate dispersion relation is excellent
to describe the ``hidden'' mode. This dispersion relation is a
generalization of the one obtained by Nelson (1976) without
shear and self-gravity, which was:

\[\tomega^2 = \nu_g^2+\frac{\tomega^2}
{\tomega^2-4\Omega^2}a_g^2q^2\]

For a ratio $a_g/a_*$ of about one fifth, we find a radial wavelength of 
about one kiloparsec, which might be varied by a factor $~2$ by varying 
the parameters.
This is the order of magnitude of the wavelength of corrugations
observed in our Galaxy and in some other edge on spirals (Quiroga  \etal 
1977,
Florido \etal 1991). Since these corrugations are observed in the very 
young stellar population, \ie are assumed to trace the motion of the gas 
disk, we consider that this ``hidden'' mode is a very good candidate to 
explain them. 

One must note that the halo contribution 
to the vertical oscillation frequency should also be included in the 
approximate dispersion relation (see \eg Toomre, 1983). 
This contribution depends only on the 
midplane density, so that for a given surface density of the halo it also 
depends on its scale height - of which nothing is known. Let us  
assume that the halo is a flattened disk in hydrostatic equilibrium. Let 
us also assume that its 
density dominates the local gravity, so that the stellar disk scale height 
is:
\[H_*\sim \frac{\rho_*}{\rho_H}\frac{a_*}{\Omega}\]
where the subscript $H$ notes halo quantities, while
\[H_H\sim \frac{a_H}{\Omega}\ .\]
Then one easily finds that the Toomre parameters of the stars and the halo 
are in the ratio:
\[\frac{Q_*}{Q_H}\sim\biggl(\frac{\rho_*}{\rho_H}\biggr)^2 \]
Thus, if both the halo and the stars have Toomre parameters not too 
different from one, their midplane densities and their contributions to 
$\nu_g$ must be comparable, so that our estimate of the corrugation 
wavelength remains valid -- though this suffers the same uncertainties as 
any estimate concerning the halo.

\section{Amplification of $m>1$ warps} In this section we briefly show 
that $m>1$ warps can be amplified, through a weak form of the Swing 
mechanism which amplifies spiral waves or modes. The $m=1$ warp, which is 
of course the most interesting since it is the one observed in most 
galactic disks, is not concerned since it has been shown by Hunter and 
Toomre (1969) that it always has a positive energy, while spiral waves are 
amplified by exchanging energy between negative energy waves inside 
corotation and positive energy ones outside corotation. This is of course 
connected with the well-known fact that the
shearing sheet model is not adapted to describe the $m=1$ mode, for which 
the effects of the cylindrical geometry cannot be neglected. 
\toutnouveau{(this is different from the mechanism studied by Bertin and
Mark (1980), who investigated the possibility of self-excited bending
modes in galactic disks through the exchange of angular momentum between
the disk and a slow bulge-halo component, with a ``quantum condition'' between
the center of the disk and the corotation.)}

\begin{figure}
\psfig{file=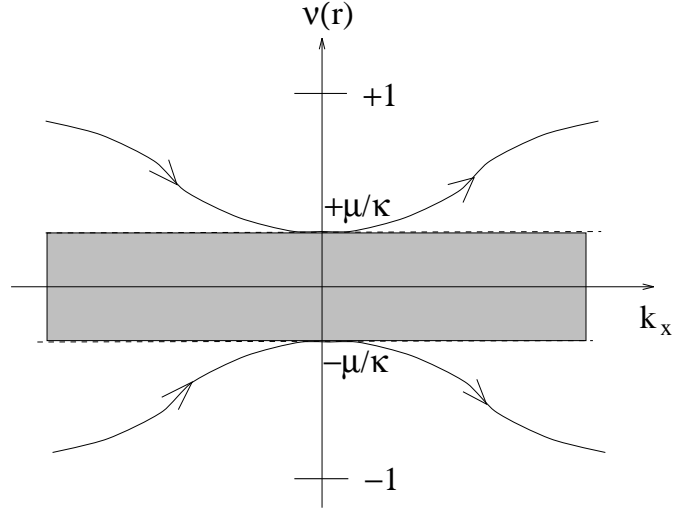,width=\columnwidth}
\caption{\label{fig:prop}
The propagation properties of bending waves are summarized in this
diagram. As in the previous figures, $\nu$ represents $\tomega/\kappa$.
It is a monotonically increasing function of $r$. See text for explanations.}
\end{figure}

\neuf{We have summarized on figure~\ref{fig:prop} the propagation properties
of bending waves. As for spiral waves, we have a forbidden band
around corotation, imposed by the vertical resonances at $+\mu$
and $-\mu$. In the case of spiral waves, the thickness of this
forbidden band tends to zero as the Toomre parameter~$Q$ gets closer
to unity. In the case of bending waves, this thickness tends to zero
as the rotation curve tends to be flat (since $\mu^2=2\Omega^2-\kappa^2$).
The sign of the group velocity, which is the same as the sign of
$\partial\nu/\partial k_x$, allows one to give the direction of propagation
of these bending waves. 
In our approximation (WKB and the perturbative derivation of the
compressibility term), the latter term becomes infinite at the 
Lindblad resonances, so that we cannot conclude here on the
exact properties of the warp close to these resonances.
We reserve a more detailed study for future work, but we have
checked that this does not affect the results presented here.}

Our motive for the investigation of non-WKB effects 
is that the dispersion relation of warps 
is similar in structure
to that of spiral waves, with a
positive rather than negative self-gravity term. It is thus analogous to 
the dispersion relation for spiral waves in a disk embedded in a strong 
vertical magnetic field, as derived by Tagger \etal (1990). They showed 
that these waves were still subject to a weak form of the Swing 
amplification mechanism, associated mathematically with the presence of 
the square root ($q=(k_x^2+k_y^2)^{1/2}$) in the self-gravity term. 
Physically, this corresponds to the long-range action of gravity or of 
magnetic stresses. Thus we can expect the same mechanism to apply to 
warps. 

However this amplification takes place when $k_x$ is of the order of 
$k_y$, so that the WKB approximation cannot be used straightforwardly. 
Although we might use the method of Pellat \etal (1990)
for an analytical derivation, we present here a full numerical solution 
which allows us to derive the result directly. We stay in the shearing 
sheet and use the formalism developed by various authors (Goldreich and 
Lynden-Bell, 1965; Toomre, {\nouveau{1964}}; 
Drury, 1980; Lin and Thurstans, 1984), 
reviewed in Tagger \etal (1994), to compute the amplification. We refer to 
these works for a complete description of the formalism, where one uses 
the fact that because of the shearing motions $k_x$ changes with time as 
$k_x^0-2Ak_yt$. Thus the $\tomega$ term obtained in the WKB approximation, 
is replaced in the Euler equation by derivatives: \[ 
\tomega\rightarrow\frac{\partial}{\partial t} 
-2Ak_y{\partial\over{\partial k_x}}\]
This in turn can be,
through a change of variables, transformed to a single derivative over 
$k_x$. There results a set of differential equations which, in the 
tightly-wound limit ($k_x\gg k_y$), reduce to the WKB result while at low 
$k_x$ they give rise to transient phenomena and amplification. We write 
and solve numerically these equations.

At large (positive or negative) $k_x$ one recovers the WKB results, while 
amplification can occur at $k_x\sim k_y$. The amplification can be 
described in the
following manner: one can combine the equations and find at large $k_x$ 
WKB solutions, varying as %
\begin{equation}
\label{eqn:WKB}
M(k_x)=P^{-1/4}\exp\left(\pm i\int_0^{k_x}P^{1/2}(k_x')dk_x'\right) 
\end{equation}
where $M$ is any perturbed quantity and $P(k_x)$ a quantity which appears 
in equations as the square of an ``instantaneous'' frequency. These 
solutions correspond to leading and trailing waves respectively for $k_x$ 
negative or positive, propagating inside or beyond corotation respectively 
for the + and - signs in the exponent. The leading waves propagate toward 
corotation, and trailing waves away from it.

Thus let us start at $k_x\rightarrow -\infty$ (\ie leading waves) with a 
``pure'' solution (\ie with the $+$ sign in the exponent), corresponding 
to a wave
traveling inside and toward corotation. The numerical integration over 
$k_x$ corresponds to following the wave as it approaches corotation, and 
then is reflected when $k_x$ becomes positive. But in the meantime the WKB 
approximation has lost its validity while $k_x\sim k_y$, so that the 
``pure'' solution has lost its identity. Thus in general, when the WKB
approximation becomes valid again (at large and positive $k_x$), we should 
obtain a {\it mixture\/}¥ of the two solutions, with the $+$ and $-$ sign 
in the exponent: this means a mixture of trailing waves traveling inside 
and outside corotation, and away from it: as the initial leading wave is 
reflected it also ``tunnels'' through the corotation region, and generates 
an outgoing trailing wave. This is the usual process of the Swing 
amplification mechanism for spiral waves; in that case the complex physics 
of this mechanism implies that the reflected wave has a larger amplitude 
than the initial one, because waves have negative energy inside corotation 
and positive energy outside. Thus that by conservation of energy the 
emission of the positive energy wave outside corotation means an 
amplification of the negative energy one. 

In the case of warps the mechanism is exactly similar; however the 
repulsive, rather than attractive, force between density perturbations on 
either side of corotation makes it much less efficient than in the case of 
spirals: one can obtain an amplification by a hundred in the most extreme 
case for spirals, while Tagger \etal (1990) find only a factor 1.05 (and a 
relative amplitude .3 for the transmitted wave) in the best case, when 
they study magnetized disks with a dispersion relation mathematically 
similar to that of warps. 

\begin{figure}
\psfig{file=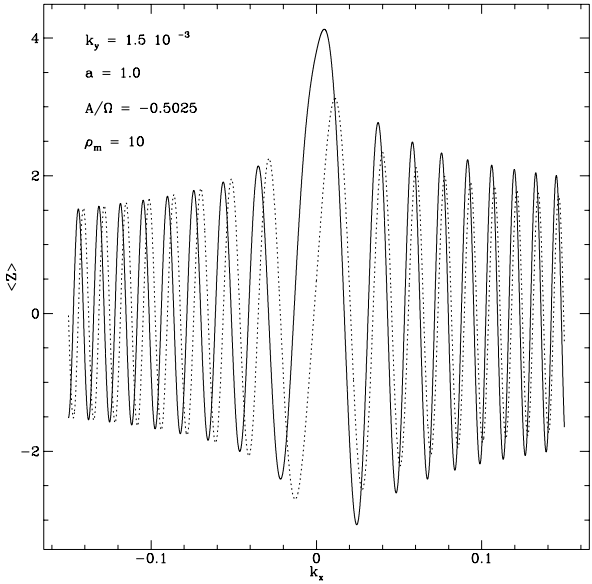,width=\columnwidth} \caption
{\label{fig:ampli}
\scriptsize
In this figure we show the mean deviation of the warp from the galactic 
plane, i.e. $\int_0^{+\infty}\rho_0(z)s(z)dz / (\Sigma/2)$, solved from 
the full set of differential equations over $k_x$. 
\neuf{The parameters of the simulation are indicated in the diagram.
We have set $4\pi G=1$ and $\Omega=1$. The units in $Z$ are arbitrary,
since the perturbation is linear}. The vertical profile of $\rho$ is
imposed through the use of the consistency equation (9) 
by the choice of the parameters $a$, $\rho_m$ and $A/\Omega$.
Here, the high value of $\rho_m$ (in our
units) and the value of $A/\Omega$ close to $-1/2$ give a disk dominated
by the local gravity, \ie with a vertical profile close to $1/\cosh^2(z/H)$.
This can be easily checked with the values of the parameters $\mu/\kappa$
and $Q$, which appear to be respectively, in this case, $7.1\times 10^{-2}$ and
$0.63$ (just above the stability limit for this type of moderately thick
disk).
The solid line 
represents the real part, and the dotted one the imaginary part. For $k_x 
< -k_y$, the solution remains WKB, with a decreasing frequency and an 
increasing amplitude (respectively due to the behavior of $\tomega$ with 
$q$ and to the $P^{-1/4}$ term in the WKB solution). Since we have only 
one wave propagating in this region, real and imaginary parts are in 
quadrature.
After crossing the region where $k_x\sim k_y$, the real and imaginary 
parts are no longer in quadrature, because the solution now projects onto 
both WKB solutions, and we see that total amplitude has changed.
}
\end{figure}

\begin{figure}
\psfig{file=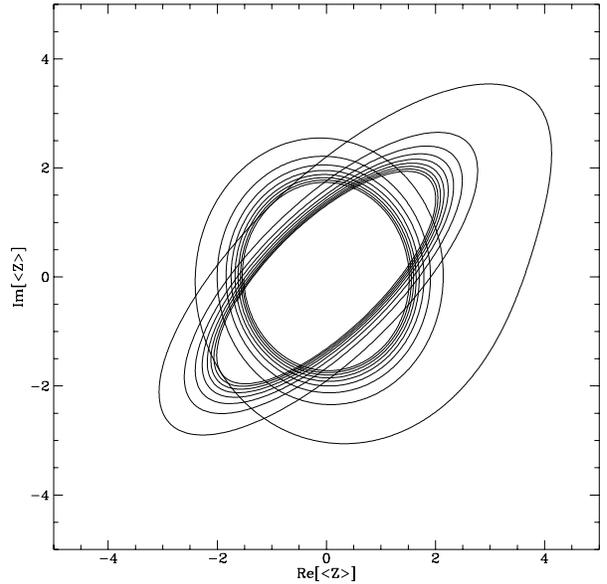,width=\columnwidth} \caption
{\label{fig:ellipse}
\scriptsize
In this figure we plot the imaginary part of the warp deviation versus its 
real part, for the same simulation as figure (7). The ``trajectory'' first 
follows circles of slowly increasing radius, as the ``pure'' leading WKB 
solution travels toward corotation. When $k_x$ has become positive, the 
trajectory becomes elliptic, because the solution is now a mixture of 
perturbations varying as $\exp\pm i\int P^{1/2}dk_x'$. The measure of the 
ellipticity gives the ratio of the transmitted to the reflected wave. }
\end{figure}

Here we obtain similar results, summarized in figures~(\ref{fig:ampli})
 and~(\ref{fig:ellipse}). This 
amplification is so weak that it is negligible for a single wave packet 
traveling in the disk. One might consider building a normal mode, formed 
as in the case
of spiral when the reflected trailing wave is reflected back at the 
galactic center to become a leading one traveling back toward corotation; 
then, with an amplification by a factor $1.05$, and assuming that the wave 
suffers no dissipation,
the $e$-folding time would be about 20 times the duration of this cycle, 
which is at best a few rotation times. This means that such modes could 
not reach a sizable amplitude in a galactic disk over a Hubble time. On 
the other hand, in accretion disks where the possible number of rotation 
periods is much larger, and still neglecting all possible sources of 
dissipation, this might provide a way of maintaining long-lived warp modes. 

We wish to note that this result has also been obtained independently by 
J. Goodman 
(private communication).

\section{Discussion}
We have investigated the dispersion relation of warps in a moderately 
thick disk. \no{We have found that the propagation of warps
is described by the dispersion relation~(\ref{eqn:RDM}),
with the following 
hypothesis:}
\begin{itemize}
\item{We work in the shearing sheet model;} \item{The disk is consistent 
(with our definition, see the corresponding section), which physically 
means that the disk must be moderately thick: it is geometrically thin 
but the ratio of the sound time through the disk to the rotation time is 
arbitrary;} \item{The warp wavelength 
is larger than the disk thickness;} \item{The disk is isothermal;}
\item{the waves are WKB, i.e. tightly wound.} \end{itemize}

The corrective term is due to the effect of compressibility, acting when 
the sound time through the disk is not much smaller than the rotation time.

The associated horizontal motions are important because they allow one to 
consider the possibility of a non-linear mechanism to continuously excite 
the warps: indeed the beat wave of two $m=1$ warps (or of a warp with 
itself) is an $m=2$ perturbation, whose potential and horizontal motion 
are even in $z$; thus they have the wavenumber and parity of a spiral. 
This means that a spiral can interact non-linearly with warps and excite 
them, in the same manner that it can do with other spirals. Non-linear 
coupling of bars and spirals has been studied by Tagger \etal, 1987, and 
Sygnet \etal, 1988. This mechanism was found very efficient and explained 
the behavior of numerical simulations of galactic spiral; in a forthcoming 
paper we will consider the possibility and the efficiency with which 
non-linear coupling of warps and spirals might feed warps from the energy 
and angular momentum carried by the spiral wave to the outer parts of the 
galactic disk. 

\new{Another potentially important effect of the compressional term relates to
the possible existence of warp modes, \ie standing wave structures, in the
disk. Classically, the Hunter-Toomre criterion shows that modes can exist
only if the integral $\int q\ dr$ is finite, and that with the thin-disk
dispersion relation this can be written as $\int dr/\Sigma$ finite - a
condition that is difficult to fulfill in realistic disk models. With the
new compressional term, this connection between the two integrals
disappears, so that one might have $\int q\ dr$ finite (\ie a discrete mode
spectrum) even when $\int dr/\Sigma$ is not bounded\footnote{As this 
paper was being rewritten following referee's comments, 
we learned that Sellwood (in preparation), using a modification 
of the dispersion relation from Toomre (1966) has reached similar 
conclusions and indeed found standing warp modes 
in numerical simulations.}.}

We have also investigated the self-amplification of $m>1$ warps by a weak 
form of the SWING mechanism. We have found an amplification which is too 
weak to significantly amplify warps in galactic disks, but might be 
considered in accretion disks.

\section{ Acknowledgments}
We wish to thank C. Pichon for rich and helpful discussions in the course 
of this work. We also thank the referee A. Romeo for very detailed
discussions and advice which have considerably enriched this paper.

\section{ Appendix: derivation of the dispersion relation}
\no{We derive hereafter the relation dispersion of warps in the
monofluid case under the hypothesis mentioned in the main text.}

Let N be:
\begin{equation}
\label{eqn:step0}
N \equiv \int_0^{+\infty}\alpha(z)dz
= \int_0^{+\infty}\frac{\rho_0}{\tomega}\biggl(a^2\frac {\partial 
s}{\partial z}+\frac{\partial\phi}{\partial z}\biggr)dz \end{equation}

where use of equation (\ref{eqn:ds})  has been made.
Using the unperturbed hydrostatic 
equilibrium  and the expression of $s$ given by
equation~(\ref{eqn:da}) we rewrite $N$ as: 

\[N = -\frac{S}{\tomega^2}\int_0^{+\infty}\frac{\partial\phi_0}{\partial 
z} \frac{\partial\alpha}{\partial z}dz + \frac{1}{\tomega} 
\biggl(\frac{a^2q^2S}{\tomega^2-\kappa^2}+1\biggr)\int_0^{+\infty} 
\rho_0\frac{\partial\phi}{\partial z}dz\] 

where we have defined
\begin{equation}
\label{eqn:S}
S = \frac{\tomega^2-\kappa^2}{\tomega^2-\kappa^2-a^2q^2} \end{equation}

The $a^2 q^2$ term in the denominator is the source of the new effect we 
introduce. It clearly represents the effect on $s$ of the compressibility 
associated with the horizontal motions. 

Integrating by parts the first term, we use the Poisson equation and the 
consistency hypothesis (i.e. that
$\Delta_r\phi_0$ does not depend on $z$), and get: 

\begin{equation}
\label{eqn:interm}
N = \frac{4\pi GS}{\tomega^2}\int_0^{+\infty}\rho_0\alpha dz - 
\frac{S\Delta_r\phi_0}{\tomega^2}N + \frac{S}{\tomega} 
\int_0^{+\infty}\rho_0\frac{\partial\phi}{\partial z}dz \end{equation}

Where we have made use of the parity properties of the warp ($s=0$ and 
$\phi=0$ at $z=0$). 

We see here that the consistency hypothesis is the key to our derivation, 
since it has allowed us to extract $\Delta_r\phi_0$ from the integral, in 
the second term of the right-hand side.

We write the solution of the Poisson equation (Equation 7) as: \[\phi = 
-e^{qz}\int_z^{+\infty}\frac{4\pi G\rho_0(z')s(z')}{2q}e^{-qz'}dz'\] 

\[-e^{-qz}\int_{-\infty}^z\frac{4\pi G\rho_0(z')s(z')}{2q}e^{qz'}dz'\] 

(one easily checks that this is the solution which verifies the boundary 
conditions at $z=0$ and at $z\rightarrow \infty$).

Since $\rho_0$ vanishes beyond a vertical scale $H$ we can expand the 
exponential to first order in $qH$; this gives, after
straightforward computations using the continuity
equation~(\ref{eqn:da}):
\[\int_0^{+\infty}\rho_0\frac{\partial\phi}{\partial z}dz = -\frac{4\pi 
G}{\tomega}\int_0^{+\infty}\!\rho_0\alpha dz +\] 

\[\frac{4\pi Gq}{\tomega}\int_0^{+\infty}\!\rho_0(z)dz 
\int_0^{+\infty}\!\!\alpha dz + O_{grav}[(qH)^2]\] 

where $O_{grav}[(qH)^2]$ means that this gravitational contribution 
also 
contains a term of second order in $qH$
which we have not evaluated, truncating the expansion of the exponentials 
to first order. An analytical computation of this can be performed if one 
assumes that the motion in the warp is vertical with a vanishing 
divergence, and if one has an analytical expression (\eg a gaussian) for 
the vertical equilibrium density profile. In these conditions one finds 
that the dispersion relation is modified by replacing the surface density 
$\Sigma$ by an apparent surface density $\Sigma '= \Sigma (1-\lambda qH)$, 
where $\lambda$ is a constant equal to $(2\pi)^{-1/2}$ in the gaussian 
case. Thus this term affects the dispersion relation only in the limit $qH 
\sim 1$, whereas the compressional term we have derived can play a role 
for an arbitrarily small value of $qH$. The computation of this gravity term could be 
well approximated by the use of
``softened gravity'' models (Erikson~1974, Athanassoula~1984,
Romeo~1994) where one artificially truncates 
the $(r-r')^{-1}$ divergence of the potential to take into account the 
geometric effect of the finite disk thickness. Thus hereafter we neglect 
this second-order gravitational term.

Substituting our result and the value of $S$ in equation 
(\ref{eqn:interm}), and dividing by $N$, we find the dispersion relation:

\[\tomega^2 = -\Delta_r\phi_0 + 2\pi G \Sigma q + 
\frac{\tomega^2}{\tomega^2-\kappa^2}a^2q^2 + O_{grav}[(qH)^2].\]


\begin{thebibliography}{}

\bibitem{}
Athanassoula, E., 1984, Phys. Rep. 114, 319 

\bibitem{}
Bertin, G., and Casertano, S., 1982,  A\&A 106, 274

\bibitem{} 
Bertin, G., and Mark, J. W.-K., 1980, A\&A 88, 289

\bibitem{}
Binney, J., 1992, ARA\&A 30, 51

\bibitem{}
Binney, J. and Tremaine, S., 1987, Galactic Dynamics, pp. 406 to 416, 
Princeton University Press 

\bibitem{}
Drury, L. O'C., 1980,  MNRAS 193 337 

\bibitem{}
Dubinski, J. and Kuijken, K., 1995,  ApJ 442, 492 

\bibitem{}
Erikson, 1974, {Ph.D. Thesis}, MIT

\bibitem{}
Florido, E., Battaner, E., Prieto, M., Mediavilla, E., and
Sanchez-Saavedra, M.L.,
1991,  MNRAS 251, 193

\bibitem{}
Goldreich, P., and Lynden-Bell, D., 1965, MNRAS 130, 125 

\bibitem{}
Hunter, C., and Toomre, A., 1969, ApJ 155, 747 

\bibitem{}
Lin, C.C. and Thurstans, R.P., 1984, {\it in} Proc. of a course on Plasma 
Astrophysics, Varenna, Italy, ESA S.P. 207 (Noordwijk, Holland), 121 

\bibitem{}
Nelson, A.H., 1976, MNRAS 177, 265 

\bibitem{}
Nelson, A.H., 1981, MNRAS 196, 557 

\bibitem{}
Papaloizou, J.C.B., Lin, D.N.C., 1995, ApJ 438, 841 

\bibitem{}
Pellat, R., Tagger, M., Sygnet, J.F., 1990, A\&A 231, 
347

\bibitem{}
Quiroga, R.J., Schlosser, W., 1977, A\&A 57, 455

\bibitem{}
Romeo, A.B., 1992, MNRAS 256, 307

\bibitem{}
Romeo, A.B., 1994, A.\&A 286, 799

\bibitem{}
Shu, F.H., 1968, Ph.D. thesis, Harvard University.

\bibitem{}
Sparke, L.S., 1984, ApJ 280, 117 

\bibitem{}
Sparke, L.S., Casertano, S., 1988, MNRAS 234, 873 

\bibitem{}
Sygnet, J.F., Tagger, M., Athanassoula, E., Pellat, R., 1988,  
MNRAS 232, 733

\bibitem{}
Tagger, M., Sygnet, J.F., Athanassoula, E., Pellat, R., 1987, ApJ 
 318, L43

\bibitem{}
Tagger M., Henriksen, R.N., Sygnet, J.-F., Pellat R., 1990, ApJ 
353, 654 

\bibitem{}
Tagger, M., Sygnet, J.-F., Pellat, R., 1994, in N-body problems and 
gravitational dynamics, proceedings of a Meeting held in Aussois, F. 
Combes and E. Athanassoula ed., p.~55 

\bibitem{}
Terquem, C., 1993, Ph.D. Thesis, Universit\'e Joseph Fourier de Grenoble

\bibitem{}
Toomre, A., 1964, ApJ 139, 1217

\bibitem{}
Toomre, A., 1966, in Proceedings of a summer study program in Geophysical Fluid Dynamics, Woods Hole Oceanographic Institution, 1966

\bibitem{}
Toomre, A., 1983, in Internal Kinematics and Dynamics of Galaxies, IAU Symp.~100,
E. Athanassoula ed., 177

\bibitem{}
Vandervoort, P.O., 1970, ApJ 161, 87


\end{thebibliography}
\end{document}